\documentclass[prl,reprint,aps,superscriptaddress,longbibliography]{revtex4-2}

\usepackage{amsmath}
\usepackage{amssymb}
\usepackage[varbb,varg]{newtx} % this replicates APS publication font
\usepackage{microtype}
\usepackage{graphicx}% Include figure files
\usepackage{dcolumn}% Align table columns on decimal point
\usepackage{bm}% bold math
\usepackage{braket}
\usepackage{feynmp}
\usepackage{graphics}
\usepackage{enumerate}
\usepackage[dvipsnames]{xcolor}

\usepackage{comment}
\usepackage[normalem]{ulem} % sout 
\usepackage{hyperref}
\hypersetup{colorlinks=true, linkcolor=NavyBlue, urlcolor=NavyBlue, citecolor=NavyBlue} % this replicates APS publication appearance
\usepackage[capitalise]{cleveref}

\everymath{\displaystyle}

\newcommand{\bmG}{{\bm G}}
\newcommand{\lB}{\ell} 
\newcommand{\alphp}[2]{\alpha^{#1}_{#2}}

\renewcommand{\Re}{\operatorname{Re}}

\begin{document}

\title{Theory of magnetoroton bands in moiré materials}

\author{Bishoy M. Kousa}
\thanks{These authors contributed equally to this work}
\affiliation{Department of Physics, The University of Texas at Austin, Austin, Texas 78712, USA}

\author{Nicol\'as Morales-Dur\'an}
\thanks{These authors contributed equally to this work}
%\thanks{nmoralesduran@flatironinstitute.org}
\affiliation{Center for Computational Quantum Physics, Flatiron Institute, New York, New York 10010, USA}
\affiliation{Department of Physics, The University of Texas at Austin, Austin, Texas 78712, USA}

\author{Tobias M. R. Wolf}
\affiliation{Department of Physics, The University of Texas at Austin, Austin, Texas 78712, USA}

\author{Eslam Khalaf}
\affiliation{Department of Physics, Harvard University, Cambridge, Massachusetts 02138, USA}

\author{Allan H. MacDonald}
\affiliation{Department of Physics, The University of Texas at Austin, Austin, Texas 78712, USA}

\date{\today}

\begin{abstract}
The recent realization of Hofstadter spectra and fractional Chern insulators in moir\'e materials has introduced a new ingredient, a periodic lattice potential, to the study of quantum Hall phases. While the fractionalized states in moir\'e systems are expected to be in the same universality class as their counterparts in Landau levels, the periodic potential can have qualitative and quantitative effects on physical observables. Here, we examine how the magnetoroton collective modes of fractional quantum Hall (FQH) states 
are altered by external periodic potentials. Employing a single-mode-approximation, we derive an effective Hamiltonian for the low-energy neutral excitations expressed in terms of three-point density correlation functions, which are computed using Monte Carlo. Our analysis is applicable to FQH states in graphene with a hexagonal boron nitride (hBN) substrate and also to fractional Chern insulator (FCI) states in twisted MoTe$_2$ bilayers. We predict experimentally testable trends in the THz absorption characteristics of
FCI and FQH states and estimate the external potential strength at which 
a soft-mode phase transition occurs between FQH and charge density wave states.
\end{abstract}

\maketitle

\emph{Introduction ---}
The recent observation of fractional quantum Hall (FQH) states in the absence of a magnetic field--known as fractional Chern insulators (FCIs) \cite{Neupert_FCI,DasSarma_Sun_FCI,Wen_FCI,Sheng_FCI,Bernevig_Regnault_FCI,Bernevig_Regnault_FCI_2}-- in twisted MoTe$_2$ (tMoTe$_2$) \cite{FCI_Experiment1,FCI_Experiment2,FCI_Transport1,FCI_Transport2} and in rhombohedral graphene aligned with hexagonal boron nitride \cite{lu2024fractional} has attracted significant attention, bringing renewed excitement to studies of the interplay between topological order and lattice translational symmetry \cite{hofstadter1976energy,Claro_Wannier_hexagonal,Claro_Square,MacDonald_Square,MacDonald_Hexagonal}. Identifying the fundamental similarities and differences between moiré FCIs and their FQH counterparts is crucial for future applications of correlated topological materials. In this regard, while a substantial body of recent work has been dedicated to understanding FCI ground state properties \cite{FCI_DiXiao,FCI_Flatiron,FCI_LiangFu,KaiSun_FCI,Crepel_FCI,Dong_CFL,Reddy_PhaseDiagram,Jiabin_FCI}, the behavior of their excitations remains relatively unexplored. Fractionalized ground states in moiré systems are in the same universality class as the Laughlin state \cite{Laughlin}, so one would expect their long-wavelength, low-energy neutral excitations to be similar to the magnetorotons \cite{GMP,GMP2} of FQH systems. The energy dispersion of these collective excitations has a minimum, which suggests an
incipient soft-mode instability to a competing charge density wave (CDW) state. 
The absence of continuous translation symmetry in the FCI case can enrich the behavior of collective excitations, giving rise to qualitative differences with respect to conventional FQH states.  These differences can have experimental consequences - 
including in some circumstances an observable violation of Kohn's theorem \cite{Kohn_Theorem}. 

\begin{figure}
    \centering   %\includegraphics[width=0.4\textwidth]{fig1_Nolines2.jpg}
    \includegraphics[width=0.49\textwidth]{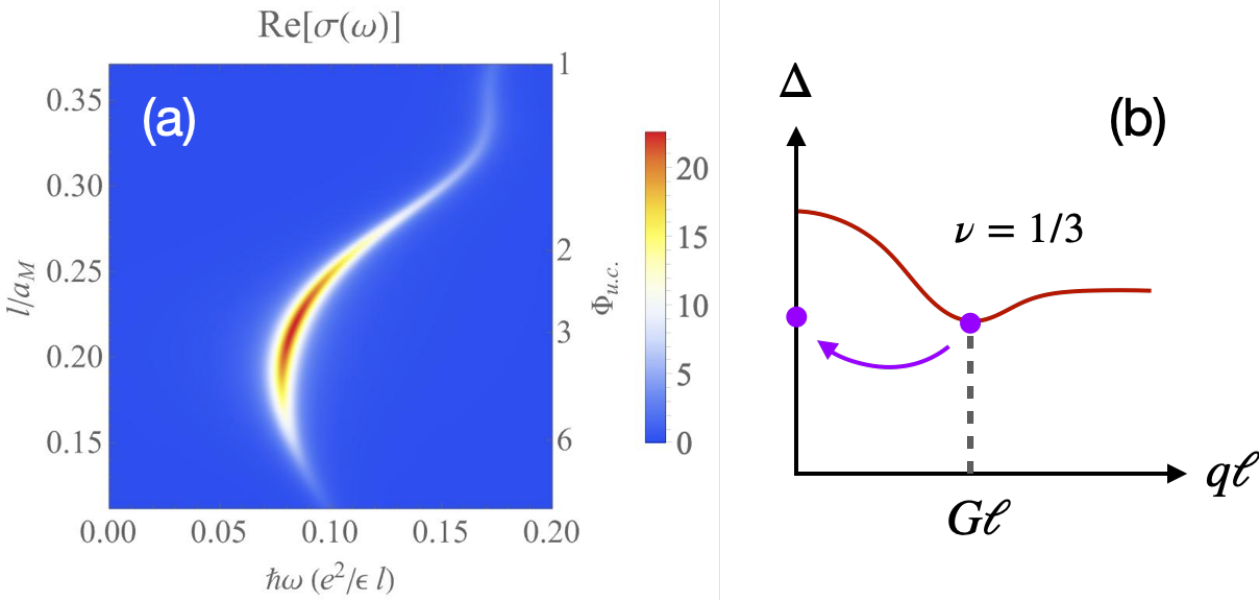}
    \caption{(a) Magnetoroton 
    band THz absorption as a function of the ratio of the magnetic length to the moir\'e period, in units of $\lambda^2 (e^2/4\hbar) \nu$ where $\lambda$ is the periodic potential strength and $\nu$ is the Landau Level filling factor [cf.~\cref{eq:conductivity_approx}].  A system of interest can be adjusted to the strong absorption regime either by varying the magnetic field or by varying the moir\'e period.  In tMoTe$_2$ fractional quantum anomalous Hall states the THz absorption is weak but can be strengthened by applying an external magnetic field. The absorption is maximized when the number of flux quanta per unit cell $\Phi_{u.c.} \approx 2.82$, which is realized at magnetic field $B \approx 15$T
    for a 30 nm moir\'e. (b) Schematic illustration of the THz conductivity mechanism.  The periodic potential reduces the energy of magnetoroton elementary excitations
    at the moir\'e Brillouin-zone corners, leading to a new ground state with reduced symmetry and observable oscillator strengths for intraband 
    collective modes.}  
    \label{fig:THz}
\end{figure}

In this Letter, we investigate one example of the consequences of 
reduced translational symmetry arising from moiré potentials in topologically ordered states. Specifically, we examine how the neutral excitations of the gapped Laughlin FQH state, the magnetorotons \cite{GMP,GMP2}, are modified when a periodic potential is added to the Hamiltonian. The external potential mixes magnetoroton 
excitations differing by a reciprocal lattice vector and folds their dispersion into the Brillouin zone defined by the moir\'e periodicity. We find that magnetoroton mixing is governed by the equal-time three-point density correlation function of the Laughlin state, which we evaluate using Monte Carlo simulations. Notably, the mixing enables moiré magnetorotons to directly couple to optical probes--in stark contrast with the case of two-dimensional electron gases in strong magnetic fields. Building on this result, we demonstrate that coupling to THz radiation is strongly enhanced when the primitive 
reciprocal lattice vectors of the periodic potential align with the undisturbed system's magnetoroton dispersion minimum, as illustrated in \cref{fig:THz}. This enhancement of the optical conductivity opens up possibilities for future spectroscopic probes of collective excitations in topologically-ordered states.

We also apply our theory to study the competition between FCI and CDW states, which has been experimentally observed at certain filling fractions in tMoTe$_2$ \cite{FCI_Experiment1,FCI_Experiment2}. As the periodic potential is made stronger, the minimum of the lowest magnetoroton band decreases to zero, signaling 
a soft-mode instability of the Laughlin state toward a competing electron Wigner crystal with Chern number $C=0$ or a hole Wigner crystal with $C=1$ \cite{HoleWC_1,HoleWC_2}. Our theory applies naturally to fractional quantum Hall states in the $N=0$ 
Landau level of graphene \cite{Kim_Graphene_LL1,Kim_Graphene_LL2} with a moiré pattern
induced either by alignment to hBN \cite{young_blg_hbn} or by a twist between 
hBN layers \cite{zhao2021universal,woods2021charge,kim2024electrostatic} in the substrate. 
It also applies to the FCIs in tMoTe$_2$ \cite{FCI_Experiment1,FCI_Experiment2,FCI_Transport1,FCI_Transport2}, since its layer-pseudospin Berry phases can be represented \cite{AdiabaticAproximation1,AdiabaticAproximation2} by an effective magnetic field with 
one flux quantum per moir\'e unit cell \footnote{Fractional quantum Hall (FQH) states are sometimes referred to 
as fractional Chern insulator (FCI) states whenever a periodic external potential is present. In this 
manuscript we reserve the term FCI for states in 
which the quantum Hall effect survives to zero magnetic field.}. Our work emphasizes qualitative differences between moiré FCI and conventional FQH states, suggests new routes for their experimental investigation and explains the wide range of twist angles over which 
FCI states are observed in tMoTe$_2$.

{\em Magnetoroton bands ---}
We consider the Landau levels generated by a constant magnetic field, $B_0$, in the presence of a periodic potential $V({\bm r})$. The cyclotron gap is given by $\hbar \omega_c=eB_0/m$, where $e$ and $m$ are the electron's charge and mass, respectively. We assume that the cyclotron gap is the largest energy scale in the problem, so that the low-energy physics is well approximated by projecting to the lowest Landau level (LLL).
When electronic interactions are added, the many-body Hamiltonian for the LLL in the presence of the periodic potential is 
\begin{align}
    H=H_{0}+V({\bm r})=\frac{1}{2A}\sum_{\bm q}V_{\bm q}\,\overline{\rho}_{\bm q}\overline{\rho}_{-\bm q}+\sum_{{\bm G}}\lambda_{{\bm G}} \,\overline{\rho}_{{\bm G}}.
    \label{MB_Hamiltonian}
\end{align}
In \cref{MB_Hamiltonian}, $V_{{\bm q}}$ is the Coulomb potential, $A$ is the area of the system, and $\overline{\rho}_{\bm{q}}$ is the density operator projected to the LLL. The reciprocal lattice vectors of the periodic potential are denoted by ${\bm G}$ and the coefficients $\lambda_{{\bm G}}$ determine the shape and strength of the potential. For concreteness, we will keep only the first shell of
reciprocal lattice vectors since higher harmonics will be suppressed by the projection to the LLL, which involves a magnetic form factor $\exp{-|{\bm G}|^2\ell^2/4}$, where $\ell$ is the magnetic length. 
We consider specifically a $C_6-$symmetric potential which is frequently relevant experimentally in 
graphene and transition metal dichalcogenide (TMD) moiré platforms. 
Under these assumptions all six first-shell Fourier coefficients are real and equal, so we denote them simply by $\lambda$.

We will focus on filling $\nu=1/3$ of the LLL, which is the simplest FQH state; our results can also be applied to $\nu=2/3$ by making a particle-hole transformation to the 
Hamiltonian \cref{MB_Hamiltonian}. In such case, the ground state of \cref{MB_Hamiltonian} when $\lambda=0$, which we denote $\ket{\Psi_0}$, is in the same universality class as the Laughlin state. When the periodic potential is turned-on, we expect a similar state to still be a good approximation for the ground state. The periodic potential will mix different Landau levels, but this effect is sub-leading in the limit of very large $\hbar\omega_c$. Within the LLL, the periodic potential couples states that differ by a reciprocal lattice vector, $\ket{\Psi_{\bm q}}$ and $\ket{\Psi_{\bm q+ {\bm G}}}$.

The low-energy neutral elementary 
excitations of \cref{MB_Hamiltonian} in the absence of the potential, $\lambda=0$, are well described by the single-mode approximation (SMA) \cite{GMP, GMP2}, \textit{i.e.} by states $\ket{\Psi_{\bm q}}$ with excitation energy $\Delta_{\bm q}$: 
\begin{align}
    \ket{\Psi_{\bm q}}=\frac{1}{\sqrt{\,N\overline{S}_{\bm q}}  }\,\overline{\rho}_{\bm q}\,\ket{\Psi_0},
    \quad 
    \Delta_{\bm q}=\frac{\overline{f}_{\bm q}}{\overline{S}_{\bm q}}.
    \label{Singlemodeapproximation}
\end{align}
Here $N$ is the number of electrons, the {\it projected oscillator strength} is
\begin{align}
    \overline{f}_{\bm q}&=\frac{1}{2N}\braket{[\overline{\rho}^{\dagger}_{\bm q},[H_0,\overline{\rho}_{\bm q}]]}_0,
\end{align}
and the {\it projected structure factor} is given by
\begin{align}
    \overline{S}_{\bm q}
    &=\frac{1}{N}\braket{\overline{\rho}^{\dagger}_{\bm q}\,\overline{\rho}_{\bm q}}_0 
    = S_{\bm q}-\left( 1-e^{-q^2\lB^2/2}\right).
    \label{eq:projected_unprojected_S}
\end{align}
In \cref{eq:projected_unprojected_S} $S_{\bm q}$ is the full structure factor, and $\langle\cdots\rangle_0\equiv\braket{\Psi_0|\cdots|\Psi_0}$. 
The SMA magnetoroton gap (\cref{Singlemodeapproximation}) can be expressed 
in terms of the static structure factor using that the projected density operators satisfy the GMP algebra $[\overline{\rho}_{\bm q},\overline{\rho}_{\bm k}]=\big( \exp(q^* k \lB^2/2)-\exp(k^* q \lB^2/2)\big) \,\overline{\rho}_{{\bm q}+{\bm k}}$, where $q=q_x+i\,q_y$ 
\footnote{The single-mode approximation is expected to be accurate for ${\bm q}$ small. For filling factor $\nu=1/3$ the SMA accurately approximates the neutral gap for $|\bm{q}|$ up to the magnetoroton minimum. 
For large $|{\bm q}|$ the SMA energy is the oscillator-strength weighted average over a continuum of neutral excitations and the neutral excitations with the lowest energy are quasi-hole-quasi-electron bound pairs.
The SMA still faithfully captures the influence of mixing between large and small $|\bm{q}|$ on the lowest energy collective modes.}. 

We would like to determine how introducing $\lambda\neq0$ changes
the spectrum of elementary excitation energies. To do so we use the SMA states as an approximate representation of the elementary single particle-hole excitation subspace.  
The external potential 
mixes SMA states related by a reciprocal lattice vector, 
but does not mix states within the Brillouin zone. In this representation the elementary excitation Hamiltonian is
\begin{align} \label{eq:matrix_elements}
    \left(h_{\bm{q}}\right)_{{\bm G},{\bm G'}} 
    &= 
    \Delta_{\bm{q}+\bm{G}}\, \delta_{{\bm G},{\bm G'}} 
    +
    \lambda \,    \frac{\braket{\overline{\rho}^{\dagger}_{{\bm q}+{\bm G}}\, \overline{\rho}_{\bm{G}-\bm{G}'}\, \overline{\rho}_{{\bm q}+{\bm G}'} }_0}{N{\sqrt{\,\overline{S}_{\bm{q}+\bm{G}} \overline{S}_{\bm{q}+\bm{G}'}}}},   
\end{align}
where $\bm{q}$ is now restricted to the Brillouin zone of the periodic potential, $\bm{G}-\bm{G}'$ belongs to the first-shell of reciprocal lattice vectors and $\Delta_{\bm{q}}$ is the magnetoroton dispersion given by \cref{Singlemodeapproximation} (see the SM \cite{Supplemental} for a derivation of this expression \footnote{The 
fractionally charged quasi-particle-quasi-hole excitations present at large ${\bm q}$ do not couple strongly to the magnetorotons, hence we do not consider them in our derivation.}.) We refer to the energies 
obtained by diagonalizing  $h_{\bm{q}}$, for all ${\bm q}$ in the Brillouin zone, as magnetoroton bands. The lowest eigenvalue of \cref{eq:matrix_elements} provides a variational estimate, within the space of SMA states, of the lowest excitation energy of the system.

\begin{figure}
    \centering   \includegraphics[width=0.48\textwidth]{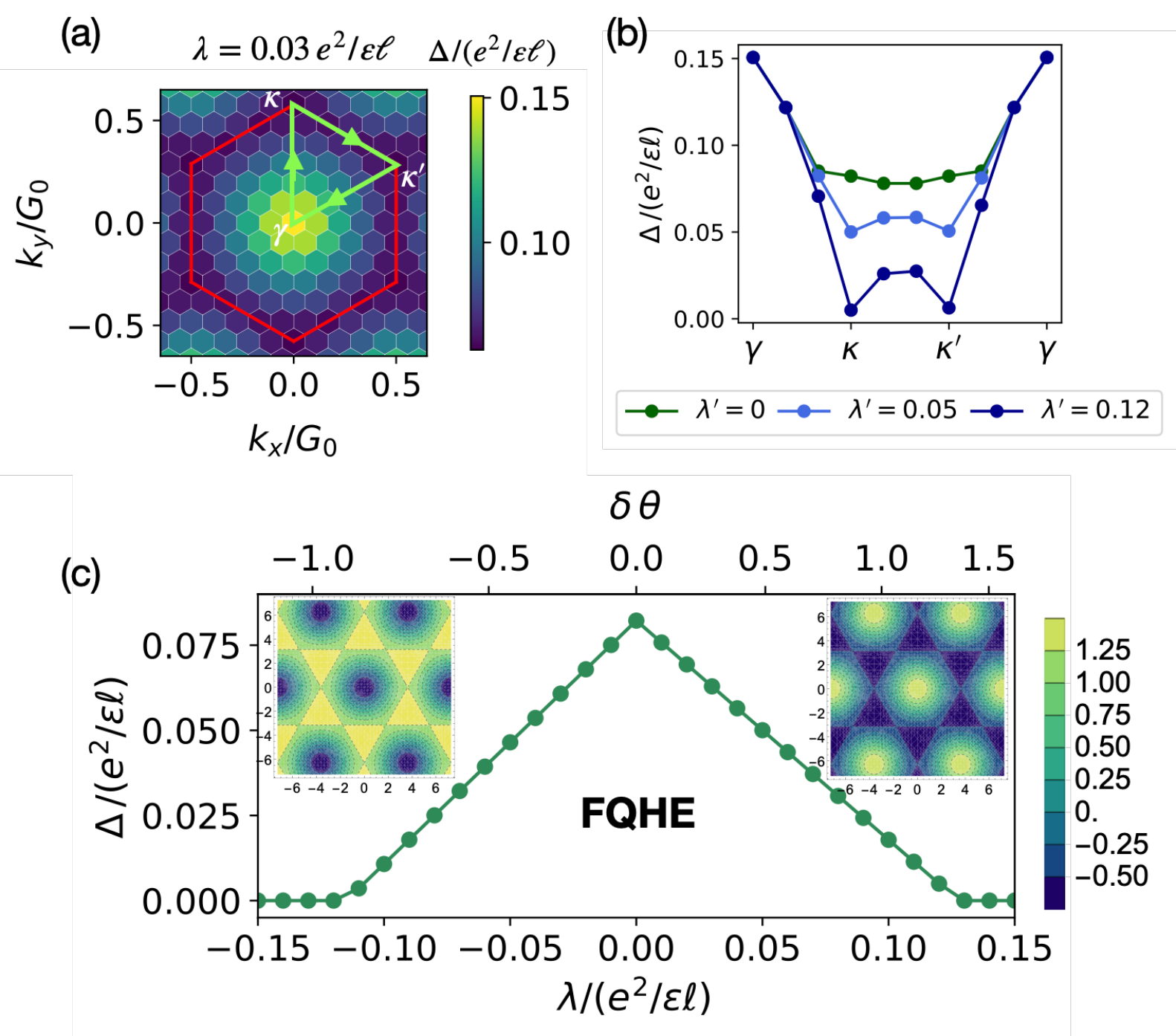}
    \caption{(a) Lowest-energy magnetoroton band for $0<\lambda<\lambda_c$, where $\lambda_c$ is the critical value for the FCI--CDW transition. The magnitude of the first-shell reciprocal lattice vectors is $G_0$. (b) Line-cut of the magnetoroton dispersion for several values of $\lambda'=\lambda/(e^2/\varepsilon\ell)$ across a path in the Brillouin zone. The magnetoroton minima are at the ${\bm \kappa}$/${\bm \kappa}^{\prime}$ points and become negative for $\lambda> \lambda_c^{+}$ or $\lambda<\lambda_c^{-}$, indicating an instability of the Laughlin-like state. (c) The magnetoroton gap $\Delta$ as a function of $\lambda$ (bottom axis) and for the tMoTe$_2$ case the deviation
    $\delta\theta$ of the twist angle $\theta$ from the magic angle $\theta_m$ value (top axis). 
    The insets indicate the symmetry of the periodic 
    potential minima for positive and negative values of $\lambda$. 
    We have truncated the reciprocal lattice to $25$ vectors and sampled the Brillouin zone with $81$ points. We have kept $2\times 10^5$ MC samples. 
    }
    \label{fig:MR_Bandstructure}
\end{figure}

The off-diagonal terms in \cref{eq:matrix_elements} involve the three-point correlation
function of the Laughlin state projected to the LLL. The latter is related to the unprojected three-point function:  
\begin{align} \label{eq:Projected_Threepoint}
    & 
    \braket{\overline{\rho}^{\dagger}_{\bm{q}_1}\overline{\rho}_{\bm{q}'}\overline{\rho}_{\bm{q}_2}}_0
    = 
    \braket{\overline{\rho^{\dagger}_{\bm{q}_1}\rho_{\bm{q}'}\rho_{\bm{q}_2}}}_0
    -
    F(
    \alphp{q_1}{q'}
    \!+\alphp{q_1}{q_2}
    \!+\alphp{-q'}{q_2}
    )
    \nonumber
    \\
    &
    -
    \delta S_{\bm{q}_2} \,
    F(\alphp{q_2}{q'})
    -
    \delta S_{\bm{q}_1-\bm{q}_2} \,
    F(\alphp{q_1}{q_2})
    -
    \delta S_{\bm{q}_1}\,
    F(\alphp{-q'}{q_2})
    ,
\end{align}
where $\bm{q}'=\bm{q}_1-\bm{q}_2$ due to momentum conservation, and we used short-hand notations $\delta S_{\bm{q}}=S_{\bm{q}}-1$, $F(\alpha)=1-e^{-\alpha/2}$ and $\alphp{q_1}{q_2}=q_1^* q_2 \lB^2$.
% %
\cref{eq:Projected_Threepoint} is a generalization of \cref{eq:projected_unprojected_S} to the three-body case (see the SM for its derivation \cite{Supplemental}.) In order to construct the Hamiltonian \cref{eq:matrix_elements} we require values for the equal-time
three-point correlation function with arbitrary momentum arguments,
which we obtain from Monte Carlo sampling of Laughlin-state position 
distributions \cite{Toby_Structurefactor}. 
Using \cref{eq:Projected_Threepoint} we then compute the projected three-point function which enters  
in the Hamiltonian \cref{eq:matrix_elements}. Typical results for the lowest magnetoroton bands are summarized in Fig.~\ref{fig:MR_Bandstructure}.

%%%%%%%%%%%%%%%%%%%%%%%%%%%%%%%%%%%%%%%%%%%%%%%%%

{\em Soft-mode instability and competing Wigner crystal states ---}
We focus first on the case of {\it one flux quantum per unit cell} of the periodic potential, which fixes a relation between the lattice length, $a$, and the magnetic length, {\it i.e.} $2\pi\ell^2=\sqrt{3}a^2/2$.
The magnetoroton band evolution with potential strength $\lambda$ in this case is 
relevant to the FCI states of tMoTe$_2$ at zero magnetic field. The continuum model for tMoTe$_2$ \cite{FengchengTopology} can be well approximated by a LLL in a weak periodic potential \cite{AdiabaticAproximation1,AdiabaticAproximation2} at a particular {\it magic} twist angle \cite{LiangFuMagic,FCI_Flatiron}, that we will denote $\theta_m$. Therefore the effective value of $\lambda$ is related to the deviation of the model parameters from magic angle values. \cref{fig:MR_Bandstructure}(a) shows the lowest magnetoroton band for $\lambda/(e^2/\varepsilon \ell) = 0.03$ and \cref{fig:MR_Bandstructure}(b) displays line traces of the magnetoroton bands across the indicated path in the Brillouin zone, for several values of the potential strength. Results for higher magnetoroton bands are presented in the SM \cite{Supplemental}. In the absence of the periodic potential the magnetoroton minimum is located near the 
centers of the edges of the Brillouin zone \cite{Ajit_MRMminimum}. As $\lambda$ is turned on, the minimum moves
toward the ${\bm \kappa}/{\bm \kappa}^{\prime}$ Brillouin-zone corners, as can be seen in \cref{fig:MR_Bandstructure}(b). The value of the magnetoroton minimum decreases as $\lambda$ is increased -- eventually reaching zero at $\lambda_c$. Soft modes at ${\bm \kappa}/{\bm \kappa}^{\prime}$ 
would accompany a continuous phase transition to a ${\sqrt{3}\times \sqrt{3}}-$unit-cell
charge density wave. This periodicity corresponds to three flux quanta and one electron per unit cell --
the periodicity of the triangular-lattice Wigner crystal state that is expected to compete with the 
FCI state. The soft mode condensation should be taken as an indication of instability of the FCI ground state towards Wigner crystals, but ultimately we expect the transition to be first-order, not continuous. 

The magnetoroton gap $\Delta$ is understood as the energy difference between the lowest neutral excitation at the ${\bm \kappa}/{\bm \kappa}^{\prime}-$points and the Laughlin state. Fig. \ref{fig:MR_Bandstructure}(c) shows the value of the magnetoroton gap as a function of $\lambda$ and also as a function of the deviation $\delta\theta$ from the magic angle $\theta_m\approx 3.75^{\circ}$, which we estimated via the adiabatic approximation \cite{AdiabaticAproximation1} and taking $\varepsilon=20$, for a continuum model of tMoTe$_2$ \cite{FCI_DiXiao}. The non-analytic behavior of the magnetoroton gap at $\lambda=0$ is expected, since three degenerate magnetoroton states at the ${\bm \kappa}/{\bm \kappa}^{\prime}-$points are coupled by the periodic potential. It follows that the lowest energy shift at this wavevector is linear in $\lambda$ --with different slopes {\it vs.} $\lambda$ in the positive and negative $\lambda$ regimes. Using the model parameters from \cite{FCI_DiXiao} we obtain that the maximum magnetoroton gap in Fig. \ref{fig:MR_Bandstructure}(c) is $\Delta\approx 2.77$ meV, while the maximum neutral gap obtained from exact diagonalization directly on the tMoTe$_2$ continuum model is $\Delta\approx 1.94$ meV for $\nu=2/3$ and $\Delta\approx 1.75$ meV for $\nu=1/3$. This indicates that our method accurately estimates the energy scales of moiré magnetorotons (See SM for details \cite{Supplemental}.)

Among the possible competing CDW states with $\sqrt{3}\times{\sqrt 3}$ order, an electron Wigner crystal with Chern number $C=0$ (signaled experimentally by the absence of a Hall response) and a hole Wigner crystal with $C=1$ (signaled experimentally by re-entrant integer quantum Hall effects) are expected to the the most competitive ones \cite{Allan_WignerCrystals}. In the SM \cite{Supplemental} we discuss the properties of these candidate phases. The potential strength $\lambda_c$ at which the transition to a CDW occurs is larger for $\lambda>0$ (a honeycomb potential -- corresponding to a hole Wigner crystal) than for $\lambda<0$ (a triangular potential -- favoring an electron Wigner crystal), in correspondence with the electron Wigner crystal being lower in energy than the hole Wigner crystal at $\nu=1/3$ \cite{Allan_WignerCrystals}. \cref{fig:MR_Bandstructure}(c) indicates that in tMoTe$_2$ the $\nu=2/3$ FCI state will be more stable than the 
$\nu=1/3$ FCI state for twist angles larger than the 
magic angle. Our calculations also explain the large twist angle ranges over which the FCI state has been observed. 

{\em THz conductivity --} 
It is well known that in conventional FQH systems intra-LL excitations do not couple directly to light (because the projected structure factor vanishes as $q^4$ in the long wavelength limit $q\to 0$ \cite{GMP}.) This result can also be seen from Kohn's theorem, which forbids any optical activity of the Laughlin state \cite{Kohn_Theorem} in a system with continuous translational symmetry. In this section, we show that adding the periodic potential, \cref{MB_Hamiltonian}, allows the intra-band collective excitations to couple directly to light, and that this coupling is enhanced when the period of the potential is such that its reciprocal lattice vectors are close to the magnetoroton minimum.
A simple expression for the THz intra-band conductivity $\Re\sigma(\omega)$ was derived previously in Ref.~\cite{Fengcheng_MoireAssisted}, based on a perturbative 
treatment of the external potential and single-mode-approximations similar to
those employed here: 
\begin{equation}
\Re\sigma(\omega) \approx \frac{e^2}{4\hbar} \nu \sum_{\bm{G}} \lB^2|\bm{G}|^2 \frac{|\lambda_{\bm{G}}|^2}{\Delta_{\bm{G}}} \,\overline{S}_{\bm{G}}\, \delta(\hbar \omega-\Delta_{\bm{G}}),
\label{eq:conductivity_approx}
\end{equation}
where $\bm{G}$ is a first-shell reciprocal lattice vector, $\nu=n_e/n_{\mathrm{LL}}$ is the LL filling fraction for electron density $n_e=N/A$ and $n_{\mathrm{LL}}=1/(2\pi\lB^2)$. This expression for $\Re\sigma(\omega)$ can be obtained by taking the long wavelength limit of 
the corresponding approximate dynamic structure factor expression,
\begin{equation}
S(\bm{q}, \epsilon) 
\approx 
\sum_{\bm{G}}
\left|
\frac{\lambda_{\bm{G}} \braket{ \overline{\rho}_{\bm{q}+\bm{G}}^{\dagger} \overline{\rho}_{\bm{q}} \overline{\rho}_{\bm{G}} }_0}{N \Delta_{\bm{G}} \sqrt{\overline{S}_{\bm{q}+\bm{G}}}}
\right|^2 
\delta\left(\epsilon-\Delta_{\bm{q}+\bm{G}}\right).
\label{eq:dynamicS}
\end{equation}
The three-point function can be approximated by observing that since there are no dipole-allowed transitions within the 
lowest Landau level, then $\overline{\rho}_{\bm{q}}|\Psi_0\rangle = 0$ to first order in $|\bm{q}|$, yielding
\begin{align}
\frac{\braket{ \overline{\rho}_{\bm{q}+\bm{G}}^{\dagger} \overline{\rho}_{\bm{q}} \overline{\rho}_{\bm{G}}}_0}{N \overline{S}_{\bm{q}+\bm{G}}} \approx i \lB^2(\bm{q} \times \bm{G}) \cdot \hat{z}.
\label{eq:3pt_approx}
\end{align}

\begin{figure}
    \centering   \includegraphics[width=0.45\textwidth]{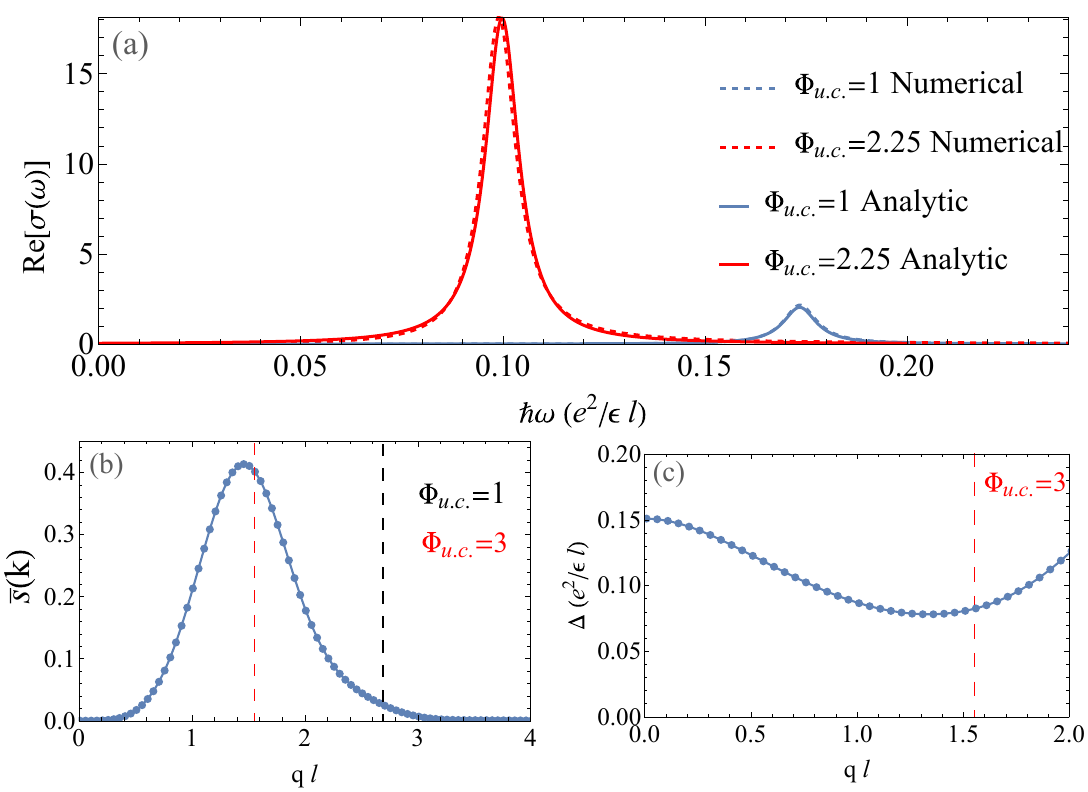}
    \caption{ 
    (a)~Comparison of optical conductivity
    %(\cref{eq:conductivity_approx} [solid lines]) curves 
    calculated using \cref{eq:conductivity_approx} [solid lines] and Monte Carlo three-point correlation functions in \cref{eq:dynamicS} [dashed lines] for different number of flux quanta piercing each unit cell $\Phi_{u.c}$. 
    (b) Projected structure factor for the Laughlin state ($\nu=1/3)$ computed using the MC fitted coefficients \cite{GMP}. The black dashed line indicates the magnitude of the moiré reciprocal lattice vector $\bm{G}$ in the tMoTe2 relevant case of one flux quantum per unit cell. The red dashed line indicates $|\bm{G}|$ for 3 flux quanta per unit cell, near the structure factor maximum (roton minimum). (c) Magnetoroton dispersion for the Laughlin state at $\nu=1/3$. The projected structure factor maximum corresponds to the roton minimum, suggesting a competing charge density wave state.    }
    \label{fig:cond}
\end{figure}

It can be seen from \cref{eq:conductivity_approx} and \cref{eq:dynamicS} that the discrete translational symmetry of the periodic potential ($\lambda_{\bm G}\neq0$) enables the FCI ground state to directly couple to optical probes--due to mixing of magnetorotons at reciprocal lattice vectors with the ground state-- and that this coupling is governed by the three-point function.
\cref{eq:conductivity_approx} was used to construct Fig.~\ref{fig:THz}, see the SM \cite{Supplemental} for details on its derivation. In \cref{fig:cond} (a) we compare line traces for typical $\sigma(\omega)$ results obtained from \cref{eq:3pt_approx} with those computed using Monte Carlo, showing excellent agreement. In the SM \cite{Supplemental}, we also compare \cref{eq:3pt_approx} with 
numerical correlation functions for different $\bm{q}$ and Monte Carlo sample sizes. Monte Carlo three-point correlation functions are the building blocks of our theory, they are also related to non-linear response 
properties \cite{peterson1967formal,gravel2007nonlinear} of quantum Hall systems
and may find further applications in Bootstrap approaches to the quantum Hall problem \cite{Boostrap_Eslam}. 

Some of us \cite{Toby_Structurefactor} have previously argued that the magnetorotons 
of tMoTe$_2$ FCI states are optically dark. To see that this conclusion is consistent with
\cref{eq:conductivity_approx}, we note that for one flux quantum per unit cell $|\bm{G}| \lB=2 (\pi /\sqrt{3})^{1/2} \approx 2.7$.  At this large wavevector 
$\overline{S}_{\bm{G}} \approx 0$ (see \cref{fig:cond} (b)), implying that the optical conductivity
remains small even for a relatively strong effective external potential.
However, one can see from \cref{eq:conductivity_approx} that if the period of the 
external potential or the magnetic field strength is tuned so
that it coincides approximately with the peak in the structure factor (and the magnetoroton minumum [c.f. Fig. \ref{fig:cond} (b-c)]),
magnetorotons will become accessible to THz spectroscopy, as illustrated in \cref{fig:THz} \footnote{In that figure we used a Lorentzian representation of the delta function $\delta(x) \approx \epsilon^2/\pi (x^2+\epsilon^2)$, $\epsilon=0.005$.}.
For tMoTe$_2$ the direction of the effective magnetic field and the applied external magnetic field 
are always the same.  It follows that the magnetic field strength needed to bring the 
magnetoroton minimum at the moir\'e reciprocal lattice vector must supply $\approx 1.7$ flux quantum per cell - requiring magnetic fields $\sim 100$ T for typical tMoTe$_2$ moir\'e periods.
We therefore conclude that the THz oscillator strength of tMoTe$_2$ magnetoroton modes 
will remain small in laboratory magnetic fields.  For the $N=0$ Landau levels of graphene aligned 
to hBN, the effective periodic potential is strongest at alignment, which yields a moir\'e period 
$a_M \sim 14$~nm.  In this case THz coupling to magnetoroton modes is maximized at $B \sim  
68$~T --still an inconveniently strong magnetic field.  We conclude that the optimal 
strategy to enable THz studies of magnetorotons is to form a long-period thBN moir\'e 
in the dielectric stack \cite{zhao2021universal,woods2021charge,kim2024electrostatic}.
For example a 30 nm moir\'e period brings the magnetic field strength at which the maximum absorption happens down to 
$B \sim 15$~T. 

\emph{Discussion ---}
Even though they are not optically accessible in the absence of external perturbation, the 
low-energy excitations of FQH systems have been observed experimentally using inelastic light scattering techniques \cite{Pinczuk_MRM1,Pinczuk_MRM2,Pinczuk_MRM3,Pinczuk_MRM4,MRM_vonKlitzing, Pinczuk_LightScattering}.  Away from the long-wavelength chiral graviton limit \cite{Haldane_Graviton1,Chiral_Gravitons}, the 
mechanism for the light scattering signal is poorly understood but thought to be disorder activated and to
give a signal proportional to the magnetoroton density of states.  
Here we have shown that the presence of a periodic potential makes the magnetoroton modes 
optically active.  At laboratory scale magnetic fields, we have shown that the optimal period for 
an external potential that is intended to make magnetoroton modes visible in THz spectroscopy is 
$\sim 30$~nm.  Periodic potentials with periods in this range have been realized using 
thBN \cite{kim2024electrostatic}, but could also be realized by periodic patterning of gate dielectrics \cite{forsythe2018band,Cano_PatternedDielectric} or by other nanolithography techniques. We believe that recent progress with van der Waals heterojunction devices and exciton spectroscopy techniques \cite{Smolenski2021WC} finally brings THz optical probes of FQH and FCI collective modes within reach. 

The single-mode approximation we employ is accurate when only one collective mode at each wavevector 
couples strongly to external potentials.  The SMA is accurate for Laughlin states at wavevectors 
up to intermediate values \cite{GMP,GMP2}, but fails at large wavevectors 
where the lowest energy excitations are fractional particle-hole excitations that have very small oscillator strengths.  In that limit the SMA energy may be viewed as an oscillator-strength weighted 
average over a continuum of excitations.  This average energy should still accurately
capture the small influence in the lowest magnetoroton band due to mixing with 
large reciprocal lattice vectors basis states.  The failure of the SMA to describe fractional
quasiparticle-quasihole excitations should therefore not limit its ability to describe the lowest 
magnetoroton band.

The framework that we have developed here to describe the collective modes of FCI and periodically 
modulated FQH states, which is based on the SMA and on the calculation of three-point equal time correlation functions, is complementary to more direct numerical approaches developed by other authors \cite{Repellin_MRM,KaiSun_MRM1,KaiSun_MRM2,ChongWang_MRM_ED,MRM_DMRG1,Hyperdeterminants,Fengcheng_MoireAssisted,Toby_Structurefactor,Debanjan_StructureFactor}. As developed here it applies only to $\nu=1/m$ Laughlin states. 
Generalizations to other FQH states, for instance those in the Jain sequence \cite{Ajit_GMP1,Ajit_GMP2},
could be attempted based on the composite fermion exciton \cite{KamillaJain1,KamillaJain2} picture of 
their collective excitations. Our conclusion that patterning on the 30 nm length scale will enable 
THz probes of FQH state collective modes, applies equally well to all those FQH states, many of 
which have rich incompletely-understood collective excitation spectra.
 
{\it Acknowledgments} -- The Flatiron Institute is a division of the Simons Foundation. 
We acknowledge HPC resources provided by the Texas Advanced Computing Center at
The University of Texas at Austin. This work was supported by a Simons Foundation Collaborative Research Grant. N.M.D. acknowleges grant NSF PHY--2309135 to the Kavli Institute for Theoretical Physics (KITP). B.M.K. was supported by NSF MRSEC DMR-2308817 through the Center for Dynamics and Control of Materials and acknowledges the hospitality of the Flatiron Institute, where part of this work has been performed.

\bibliography{refs}

%%%%%%%%%%%%%%%%%%%%%%%%%%%%%%%%%%%%%%%%%%%%%%%%%
%%%%%%%%%%%%%%%%%%%%%%%%%%%%%%%%%%%%%%%%%%%%%%%%%
%%%%%%%%%%%%%%%%%%%%%%%%%%%%%%%%%%%%%%%%%%%%%%%%%
%%%%%%%%%%%%%%%%%%%%%%%%%%%%%%%%%%%%%%%%%%%%%%%%%
%%%%%%%%%%%%%%%%%%%%%%%%%%%%%%%%%%%%%%%%%%%%%%%%%
%%%%%%%%%%%%%%%%%%%%%%%%%%%%%%%%%%%%%%%%%%%%%%%%%

\clearpage

\appendix
\onecolumngrid
\section*{Supplemental material for \\ ``Theory of magnetoroton bands in moiré materials"}

\subsection{Brillouin-zone-folded Hamiltonian}
As commented in the main text, the periodic potential will couple the ground state $\ket{\Psi_0}$, with energy $E_0$, to excitations with momentum equal to a first-shell reciprocal lattice vector, modifying the ground state energy. We will restrict the manifold of excitations to only contain the magnetoroton states $\ket{\Psi_{\bm G}}=\overline{\rho}_{\bm G}\ket{\Psi_0}/\sqrt{\,N \bar{S}_{\bm G}}$, as they accurately capture the long-wavelength limit of the neutral excitations, up to the magnetoroton minimum for $\nu=1/3$ filling factor. The first-order correction to the ground state energy, that we denote $E_0^{(1)}$, vanishes in the thermodynamic limit due to continuous magnetic translation symmetry; more precisely the momentum difference between two topological ground states in a torus is proportional to the system size and the potential considered here only includes the first-shell of reciprocal lattice vectors. The second order correction to the ground state energy is given by
\begin{align}
    E_0^{(2)}=\lambda^2\sum_{\bm G}\frac{|\braket{\Psi_{\bm G}|\overline{\rho}_{\bm G}|\Psi_0}|^2}{E_0-E_{\bm G}}= -6\,\lambda^2\frac{|\braket{\overline{\rho}_{\bm G}^{\dagger}\overline{\rho}_{\bm G}}_0|^2}{N\Delta_{\bm G}\bar{S}_{\bm G}},%=-6\lambda^2 \frac{(S_GN)^2}{N\Delta_{\bm G}S_{\bm G}}\sim \lambda^2 N,
\end{align}
where $\ket{\Psi_{\bm G}}$ is a magnetoroton excitation. Note that in the main text we are treating the magnetorotons as bosonic collective modes when we assume that the extensive energy shift $\propto N$ due to the periodic potential perturbation is shared by ground state and elementary excitation spaces. Diagonalizing \cref{eq:matrix_elements} in the main text does not account for this identical overall shift. These many-body state energy 
corrections arise from coupling between subspaces with different numbers of particle-hole excitations, are $\propto N \lambda^2$ and generically differ 
by $\sim 1 \times \lambda^2$ between ground and elementary excitation subspaces. We do not discuss these shifts further below since the dominant effect we identify is $\sim \lambda$.\\\\
Although our approach to solve for the lowest elementary excitation energies is variational and is valid to arbitrary order in the periodic potential strength $\lambda$, it is also possible to do perturbative calculations. In particular, something special happens at the six Brillouin zone corners --namely that they occur in two groups of three, with any two members of a triplet $\{\bm{\kappa}_1, \bm{\kappa}_2,\bm{\kappa}_3 \}$ connected by a reciprocal lattice vector:
\begin{align}
    \braket{\Psi_{\bm \kappa_1+\bm {G}}|\overline{\rho}_{{\bm G}}|\Psi_{\bm \kappa_2}}_0 \neq 0,
\end{align}
and $|\bm{\kappa}_2| = | \bm{\kappa}_1+\bm {G}|$. If we do such a perturbative calculation at the $\bm \kappa-$point, we must use degenerate-state perturbation theory because three excitations that have the same energy at $\lambda=0$ are coupled by the perturbation. It follows that the lowest energy shift is linear in $\lambda$ at this wavevector. The lowest excitation energy at $\bm \kappa$ must shift downward since the degenerate-state perturbation matrix is traceless.
Because the lowest eigenvalue goes down linearly at this wavevector and quadratically at other wavevectors, the minimum along the zone-boundary moves quickly to the BZ corner as $|\lambda|$ increases. \\\\
Since the dominant corrections to the magnetoroton energies due to the periodic potential are of order $\lambda$, we focus our analysis on the effect of the periodic potential on the branch low-energy neutral excitations. To obtain the Brillouin-zone-folded (BZ-folded) Hamiltonian \cref{eq:matrix_elements} in the main text, we use the normalized SMA basis \cref{Singlemodeapproximation}, i.e., 
\begin{equation}
    \braket{\Psi_{\bm q'}|\Psi_{\bm q}}=\frac{1}{N\sqrt{\,\overline{S}_{\bm q'}\, \overline{S}_{\bm q}}  }\, \braket{\Psi_0|\overline{\rho}^\dagger_{\bm q'}\overline{\rho}_{\bm q}|\Psi_0}=\delta_{\bm q', \,\bm q},
\end{equation}
where $\overline{S}_{\bm q}$ is the projected structure factor in \cref{eq:projected_unprojected_S} and we have translational symmetry in the Laughlin state. Writing the Schrödinger equation in this SMA state basis, i.e.,  
%
% \begin{equation}
$
    \ket{\Phi_{\bm q}}=\sum_{\bm G} c_{{\bm q}+{\bm G}}\ket{\Psi_{\bm {q}+{\bm G}}}
$
% \end{equation}
%
with $\bm q$ is restricted to the Brillouin zone, and taking the inner product with $\ket{\Psi_{\bm q+\bm{G'}}}$ on both sides, we find 
\begin{equation}
    \sum_{\bm G} H(\bm q)_{{\bmG}',\bm G} c_{{\bm q}+{\bm G}} =
    E\sum_{\bm G} c_{{\bm q}+{\bm G}} \braket{\Psi_{\bm {q}+{\bm G'}}|\Psi_{\bm q+{\bm G}}}=c_{{\bm q}+{\bm G}'}.
\label{eq: ham_expansion}
\end{equation}
This leads to an eigenvalue problem for the Hamiltonian matrix and the expansion coefficients, with states within the BZ decoupled. The Hamiltonian matrix elements are $H(\bm q)_{{\bmG}',\bm G}=\braket{\Psi_{\bm {q}+{\bm G}'}|H|\Psi_{\bm q+\bm{G}}}$. In the SMA basis, the $H_0$ term is diagonal:
\begin{equation}
    H_0(\bm q)_{{\bmG}',\bm G}=\braket{\Psi_{\bm {q}+{\bm G}'}|H_0|\Psi_{\bm q+\bm{G}}}=  \delta_{\bm G', \,\bm G} \braket{\Psi_{\bm {q}+{\bm G}'}|H_0|\Psi_{\bm q+\bm{G}}}=  \delta_{\bm G', \,\bm G}  \Delta(\bm{q+G}),
\end{equation}
where $\Delta({\bm q}+\bm G)$ is the unperturbed magnetoroton dispersion, while the perturbing periodic potential $V$ couples states related by a reciprocal lattice vector, i.e., 
\begin{equation}
    \braket{\Psi_{\bm {q}+{\bm G'}}|V|\Psi_{\bm q+{\bm G}}}=\lambda \sum_{\tilde{\bm G}}\braket{\Psi_{\bm {q}+{\bm G'}}|\overline{\rho}_{\tilde{\bm G}}|\Psi_{\bm q+{\bm G}}}=\lambda \sum_{\tilde{\bm G }} \delta_{\bm G',  \tilde{\bm G}+\bm G} \frac{1}{N \sqrt{\, \overline{S}_{\bm{q} \, + \bm{G}}\,\overline{S}_{\bm{q} \, + \bm{G}'}}} \braket{\Psi_0|\overline{\rho}^{\dagger}_{{\bm q}+{\bm G}'} \overline{\rho}_{\tilde{{\bm G}}} \overline{\rho}_{{\bm q}+{\bm G}} |\Psi_0}. 
\end{equation}
As stated in the main text, the problem thus reduces to diagonalizing the effective Hamiltonian 
\begin{equation}
    H(\bm q)_{{\bmG}',\bm G}=\delta_{\bm G', \,\bm G}  \Delta(\bm{q+G})+\lambda \sum_{\tilde{\bm G }} \delta_{\bm G',  \tilde{\bm G}+\bm G} \frac{1}{N \sqrt{\, \overline{S}_{\bm{q} \, + \bm{G}}\,\overline{S}_{\bm{q} \, + \bm{G}'}}} \braket{\Psi_0|\overline{\rho}^{\dagger}_{{\bm q}+{\bm G}'} \overline{\rho}_{\tilde{{\bm G}}} \overline{\rho}_{{\bm q}+{\bm G}} |\Psi_0}.
\end{equation}

% \begin{equation}
%     \braket{\Psi_{\bm {q'}+{\bm G}'}|H_0|\Psi_{\bm q+\bm{G}}}= \delta_{\bm q', \,\bm q} \delta_{\bm G', \,\bm G} \braket{\Psi_{\bm {q'}+{\bm G}'}|H_0|\Psi_{\bm q+\bm{G}}}= \delta_{\bm q', \,\bm q} \delta_{\bm G', \,\bm G}  \Delta(\bm{q+G})
% \end{equation}

%and the periodic potential $V$ couples states related by a reciprocal lattice vector

% \begin{equation}
%     \braket{\Psi_{\bm {q'}+{\bm G'}}|V|\Psi_{\bm q+{\bm G}}}=\lambda \sum_{\tilde{\bm G}}\braket{\Psi_{\bm {q'}+{\bm G'}}|\overline{\rho}_{\tilde{\bm G}}|\Psi_{\bm q+{\bm G}}}=\lambda \,\delta_{\bm q', \,\bm q}\delta_{\bm G',  \tilde{\bm G}+\bm G} \frac{1}{N \sqrt{\, \overline{S}(\bm{q} \, + \bm{G})\overline{S}(\bm{q}' \, + \bm{G}')}} \braket{\Psi_0|\overline{\rho}^{\dagger}_{{\bm q}+{\bm G}'} \overline{\rho}_{\tilde{{\bm G}}} \overline{\rho}_{{\bm q}+{\bm G}} |\Psi_0}
% \end{equation}

% \cref{eq: ham_expansion} then becomes

% \begin{equation}
%     H(\bm{q})_{\bm{G},{\bm G}'} c_{\bm q'+\bm G'}
% \end{equation}

\subsection{Three-point function projected to the LLL}
We use the framework proposed by Girvin and Jach \cite{GirvinJach} to project an operator that depends on $z$ and $z^*$ into the lowest Landau level -- by replacing $z^*\mapsto 2\lB^2\frac{\partial}{\partial z}$ -- and we use complex notation $q=q_x+i q_y$. The three-point density correlation function can be split into five terms: 

\begin{align}
    \overline{\rho}^{\dagger}_{{\bm q}+{\bm G'}}\,\overline{\rho}_{\widetilde{{\bm G}}}\overline{\rho}_{{\bm q}+{\bm G}}=& \left( \sum_{j}e^{-i(-q-G^{\prime})\lB^2\frac{\partial}{\partial z_j}}e^{{-i(-q-G^{\prime})^*}\frac{z_j}{2}}\right) \left( \sum_{n}e^{-i \tilde{G} \lB^2\frac{\partial}{\partial z_n} }e^{-i\tilde{G}^*\frac{z_n}{2}}\right) \left( \sum_{m}e^{-i(q+G)\lB^2\frac{\partial}{\partial z_m}}e^{{-i(q+G)^*}\frac{z_m}{2}}\right) \nonumber \\
    =&N\, e^{[-(q+G^{\prime})^*\tilde{G}+\tilde{G}^* (q+G)-(q+G^{\prime})^*(q+G))]\lB^2/2}\nonumber\\
    &+\sum_{j\neq n \neq m }\left(e^{-i[(-q-G^{\prime})\frac{\partial}{\partial z_j}+\tilde{G}\frac{\partial}{\partial z_n}+(q+G)\frac{\partial}{\partial z_m}]\lB^2} \right)\left(e^{-i[(-q-G^{\prime})^*\frac{z_j}{2}+\tilde{G}^*\frac{z_n}{2}+(q+G)^*\frac{z_m}{2}]} \right)\nonumber \\
    &+\sum_{j= n \neq m }\left( e^{-i(-q-G^{\prime}+\tilde{G})\lB^2\frac{\partial}{\partial z_j}}e^{-i(q+G)\lB^2\frac{\partial}{\partial z_m} }e^{-i(-q-G^{\prime}+\tilde{G})^*\frac{z_j}{2}}e^{-i(q+G)^*\frac{z_m}{2}}\right)e^{-(q+G^{\prime})^*\tilde{G}\lB^2/2}\nonumber \\
    &+ \sum_{j= m \neq n }\left(e^{-i(-q-G^{\prime}+q+G)\lB^2\frac{\partial}{\partial z_j}}e^{-i\tilde{G}\lB^2\frac{\partial}{\partial z_n}}e^{-i(-q-G^{\prime}+q+G)^*\frac{z_j}{2}}e^{-i\tilde{G}^*\frac{z_n}{2}}\right)e^{-(q+G^{\prime]})^*(q+G)\lB^2/2}\nonumber \\
    &+ \sum_{j\neq m = n }\left(e^{-i(-q-G^{\prime})\lB^2\frac{\partial}{\partial z_j}}e^{-i(\tilde{G}+q+G)\lB^2\frac{\partial}{\partial z_n}}e^{-i(-q-G^{\prime})^*\frac{z_j}{2}}e^{-i(\tilde{G}+q+G)^*\frac{z_n}{2}}\right)e^{\tilde{G}^*(q+G)\lB^2/2}.
\end{align}
The Laughlin-state expectation value of this operator is non-vanishing only when momentum is conserved (up to a reciprocal lattice vector), requiring $\tilde{\bm{G}}$ such that ${\bm G}+\tilde{{\bm G}}-{\bm G}^{\prime}=0$. We can then rewrite the last three lines using full two-point functions: 
\begin{align}
    \overline{\rho}^{\dagger}_{{\bm q}+{\bm G'}}\,\overline{\rho}_{\widetilde{{\bm G}}}\overline{\rho}_{{\bm q}+{\bm G}}=&~N\, e^{[-(q+G^{\prime})^*\tilde{G}+\tilde{G}^* (q+G)-(q+G^{\prime})^*(q+G))]\lB^2/2}\nonumber\\
    &+\sum_{j\neq n \neq m }\left(e^{-i[(-q-G^{\prime})\frac{\partial}{\partial z_j}+\tilde{G}\frac{\partial}{\partial z_n}+(q+G)\frac{\partial}{\partial z_m}]\lB^2} \right)\left(e^{-i[(-q-G^{\prime})^*\frac{z_j}{2}+\tilde{G}^*\frac{z_n}{2}+(q+G)^*\frac{z_m}{2}]} \right)\nonumber \\
    &+ \left(\overline{\rho_{{-\bm q}-{\bm G}}\rho_{{\bm q}+{\bm G}}}-N\right)e^{-(q+G^{\prime})^*\tilde{G}\lB^2/2}+\left( \overline{\rho_{-\widetilde{{\bm G}}}\rho_{\widetilde{{\bm G}}}}-N \right) e^{-(q+G^{\prime]})^*(q+G)\lB^2/2}\nonumber \\
    &+ \left(\overline{\rho_{{-\bm q}-{\bm G}^{\prime}} \rho_{{\bm q}+{\bm G}^{\prime}}} -N\right)e^{\tilde{G}^*(q+G)\lB^2/2}.
    \label{threepoint_sup1}
\end{align}
Finally, in the previous expression, the term where all three indices are different ($i\neq j\neq k$) is related to the full unprojected three-point function through 
\begin{align}
    \sum_{j\neq n \neq m }&\left(e^{-i[(-q-G^{\prime})\frac{\partial}{\partial z_j}+\tilde{G}\frac{\partial}{\partial z_n}+(q+G)\frac{\partial}{\partial z_m}]\lB^2} \right)\left(e^{-i[(-q-G^{\prime})^*\frac{z_j}{2}+\tilde{G}^*\frac{z_n}{2}+(q+G)^*\frac{z_m}{2}]} \right) \nonumber \\
    &=\left(\rho^{\dagger}_{{\bm q}+{\bm G'}}\,\rho_{\widetilde{{\bm G}}}\,\rho_{{\bm q}+{\bm G}}-N\right)-\left(\overline{\rho_{{-\bm q}-{\bm G}}\rho_{{\bm q}+{\bm G}}}-N\right)-\left( \overline{\rho_{-\widetilde{{\bm G}}}\rho_{\widetilde{{\bm G}}}}-N \right) - \left(\overline{\rho_{{-\bm q}-{\bm G}^{\prime}} \rho_{{\bm q}+{\bm G}^{\prime}}} -N\right).
    \label{threepoint_sup2}
\end{align}
Inserting \cref{threepoint_sup2} into \cref{threepoint_sup1} and dividing by $N$ leads to the projected three-point function presented in the main text, and repeated here for the reader's convenience:
\begin{align} 
    & 
    \frac{1}{N}\braket{\overline{\rho}^{\dagger}_{{\bm q}+{\bm G'}}\,\overline{\rho}_{\bm{G}''}\overline{\rho}_{{\bm q}+{\bm G}}}_0
    =
    \frac{1}{N}\braket{\overline{\rho^{\dagger}_{{\bm q}+{\bm G'}}\,\rho_{\bm{G}''}\,\rho_{{\bm q}+{\bm G}}}}_0
    -\left[ 1-e^{[-(q+G')^*G''+G^{\prime\prime*}(q+G)-(q+G')^*(q+G)]\,\frac{\lB^2}{2}}\right]
    % \nonumber 
    \\
    &
    \qquad\qquad
    -\left[ S_{{\bm q}+ {\bm G}}-1\right]\left[ 1-e^{-(q+G')^*G''\,\frac{\lB^2}{2}}\right]
    -\left[ S_{\bm{G}''}-1\right]\left[ 1-e^{-(q+G')^*(q+G)\,\frac{\lB^2}{2}}\right] 
    \nonumber
    % \\
    % %
    % &
    -\left[ S_{{\bm q}+ {\bm G}'}-1\right]\left[ 1-e^{G^{\prime\prime*}(q+G)\,\frac{\lB^2}{2}}\right].
\end{align}
The difference between products of projected and products of unprojected 
density operators originates from virtual transitions to higher Landau levels; the 
projected product vanishes in the case of a fully-filled Landau level because these are 
the only transitions available \cite{Giuliani_Book}.
\subsection{Possible competing phases}
\begin{figure}
    \centering   \includegraphics[width=0.75\textwidth]{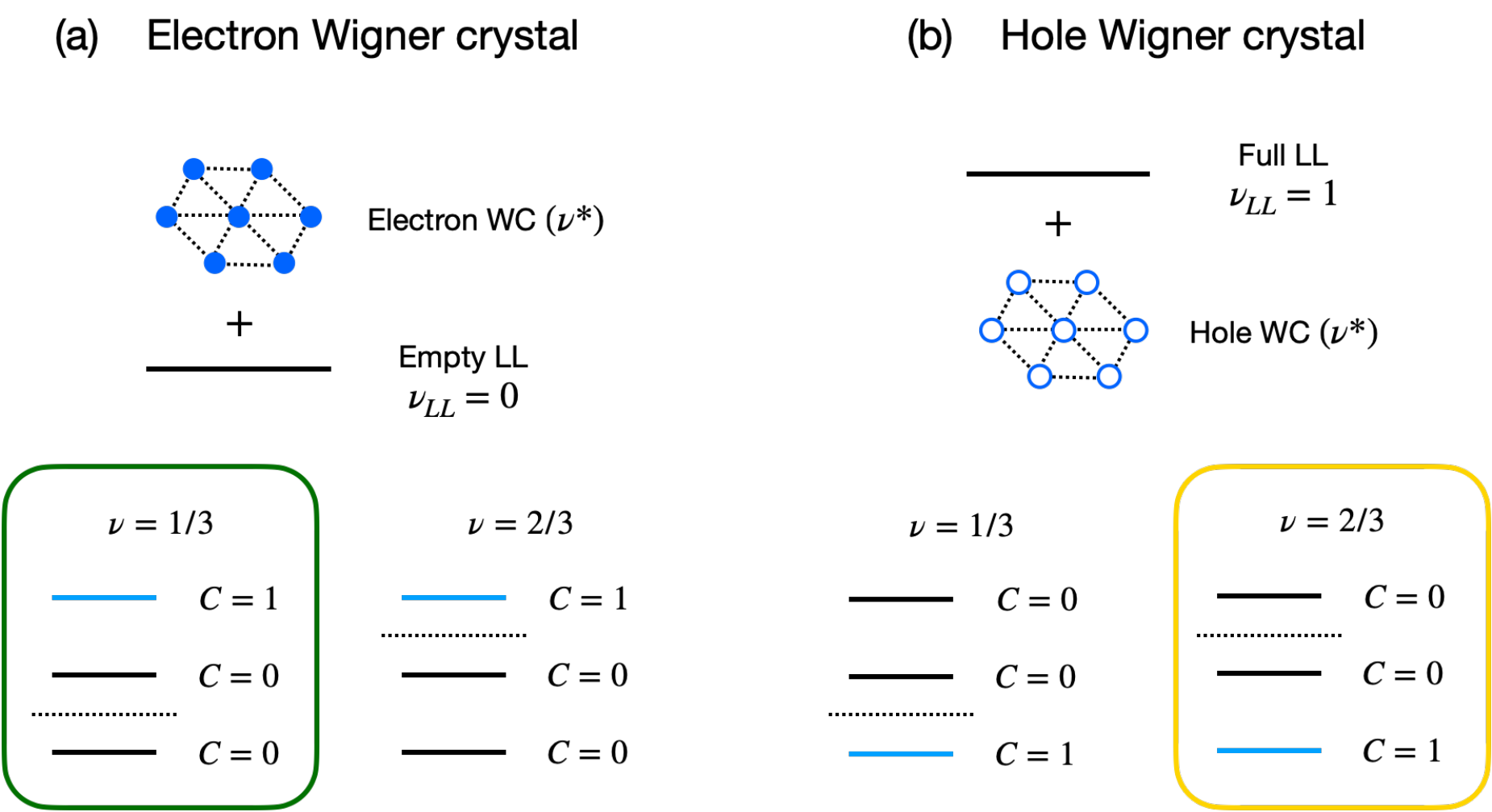}
    \caption{Schematics of two possible competing phases to the FCI state in moiré systems. (a) Electron Wigner crystal with $C=0$ and (b) Hole Wigner crystal with $C=1$. The total electronic filling fraction is $\nu=\nu_{LL}+\nu*$. The Wigner crystal of electrons or holes does not contribute to the transport properties, hence the Chern number of the phase is determined by the LL being either full or empty. The competing states can also be understood from a splitting of the LL into three sub-bands, due to the translation symmetry breaking. The dashed lines indicate the number of filled sub-bands. Previous Hartree-Fock calculations \cite{Allan_WignerCrystals} showed that, in Landau levels,  at $\nu=1/3$ the most competitive state is the electron Wigner crystal (green box), while at $\nu=2/3$ the hole Wigner crystal (yellow box) has lower energy.}
    \label{fig:CompetingPhases}
\end{figure}

For the case of one flux quantum per unit cell (applicable to twisted MoTe$_2$) the magnetoroton excitation condenses at the ${\bm \kappa}$ and ${\bm \kappa^{\prime}}$ points in the Brillouin zone, which  correspond to the wave-vector of translation-symmetry breaking to a state with a $\sqrt{3}\times{\sqrt 3}$ enlarged unit cell.\\\\
Among the possible states with $\sqrt{3}\times{\sqrt 3}$ charge order, previous studies on Landau levels (LLs) \cite{Allan_WignerCrystals} found an electron Wigner crystal and a hole Wigner crystal to be the most energetically competitive. The electron Wigner crystal forms when electrons are added to an empty Landau level and is topologically-trivial with Chern number $C=0$, as illustrated in Fig. \ref{fig:CompetingPhases}(a). In contrast, the hole Wigner crystal (signaled experimentally
by re-entrant integer quantum Hall effects) can be understood as starting from a fully-filled Landau level and removing holes from it, the gas of holes on top of the full LL forms a Wigner crystal, which does not contribute to transport, hence the Chern number of this phase is the same as for the full LL, namely $C=1$ \cite{HoleWC_1,HoleWC_2}. This state is illustrated in Fig. \ref{fig:CompetingPhases}(b).  In Landau levels, the electron
Wigner crystal at filling factor $\nu$ and hole Wigner crystal at filling factor $1-\nu$ are related by an exact particle-hole symmetry and compete equally for the ground state.\\\\
Alternatively, one could understand the translation-symmetry breaking competing phases from a band-folding perspective: Due to the $\sqrt{3}\times{\sqrt 3}$ order, the LL folds into three sub-bands whose Chern number sequence is $C=(0,0,1)$, as illustrated in Fig. 1. Mean-field calculations have revealed that at $\nu=1/3$ the most competitive phase is the $C=0$ electron Wigner crystal, while at $\nu=2/3$ the $C=1$ hole Wigner crystal is most competitive with respect to the FQH ground state \cite{Allan_WignerCrystals}. 
\subsection{Derivation of the perturbative expression of the dynamical structure in the single mode approximation}
\begin{figure}
    \centering   \includegraphics[width=0.9\textwidth]{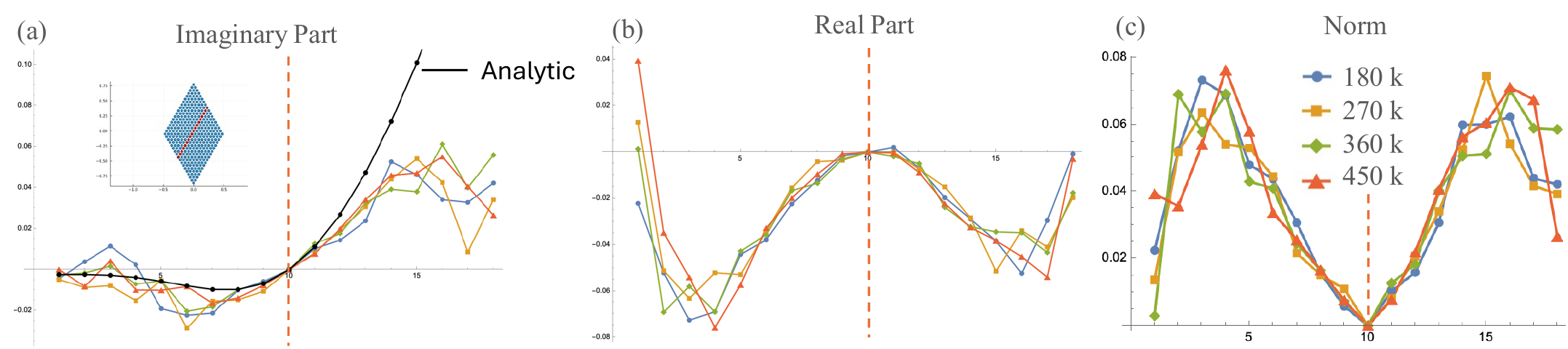}
    \caption{ Comparison the the numerical and analytical expressions of the long wavelength limit of the three point function's \cref{eq:small_q_3pt} imaginary part (a), real part (b), and norm (c), for a line cut across the BZ mesh indicated in the inset of (a). The dashed line corresponds to $\bm q =0$. The different lines are different number of MC samples shown in (c). 
}
    \label{fig:comparison}
\end{figure}

In this section, we derive the perturbative expression for the optical conductivity in the SMA. The first step is to use perturbation theory for the first order correction of the wavefunctions due to the periodic potential $\ket{\Psi_{\bm k ,m}} \approx \ket{\Psi_{\bm k ,m}^{(0)}}+\ket{\Psi_{\bm k ,m}^{(1)}}$ where 
\begin{equation}
    \ket{\Psi_{\bm k ,m}^{(1)}} = \sum_{\bm G,n} \lambda_{\bm G} \frac{\braket{\Psi_{\bm k+\bm G ,n}^{(0)}|\bar{\rho}_{\bm G}|\Psi_{\bm k ,m}^{(0)}}}{E^{(0)}_{\bm k,m}-E^{(0)}_{\bm k+ \bm G,n}} \ket{\Psi_{\bm k +\bm G,n}^{(0)}},
\end{equation}

to calculate the dynamical structure factor

\begin{equation}
S(\mathbf{q}, \epsilon)=\frac{1}{N} \sum_{m>0}  |\braket{\Psi_m|\bar{\rho}_{\bm q}|\Psi_0}|^2\delta\left[\epsilon-\left(E_m-E_0\right)\right],
\label{eq:SM_dynamical_S}
\end{equation}
where $m$ labels the excited states ($m=0$ labels the ground state). Keeping only leading order terms in $\lambda$ and noting that the unperturbed states matrix elements $\braket{\Psi_{\bm q,m}^{(0)}|\bar{\rho}_{\bm q}|\Psi_0^{(0)}}$ scale as $|\bm q|^2$ in the long wavelength limit, so they will not contribute to the optical conductivity
\begin{equation}
    \mathrm{Re} ~\sigma (\omega)= \frac{\pi e^2 N}{A} ~\mathrm{lim}_{\bm q\to 0} ~[\omega S(\bm q, \hbar \omega) /q^2],
    \label{eq:SM_cond}
\end{equation}
the only leading order term that contributes to the dynamical structure factor is 
\begin{equation}
    \braket{\Psi_{\bm k, m}^{(0)}|\bar{\rho}_{\bm q}|\Psi_0^{(1)}}=\sum_{\bm G, n}\lambda_{\bm G} 
 \frac{\braket{\Psi_{\bm G,n}^{(0)}|\bar{\rho}_{\bm G}|\Psi_{0}^{(0)}}}{E_0^{(0)}-E_{\bm G,n}^{(0)}} \braket{\Psi_{\bm q +\bm G,m}^{(0)}|\bar{\rho}_{\bm q}|\Psi_{\bm G,n}^{(0)}}\delta_{\bm k, \bm q + \bm G} 
 \label{eq:SM_S_mat_elem}
\end{equation}
Substituting Eq. \cref{eq:SM_S_mat_elem} in Eq. \cref{eq:SM_dynamical_S}, we get
%, to leading order in $\bm q$ and $\lambda$
\begin{equation}
    S(\bm q,\epsilon)=\frac{1}{N} \sum_{\bm G, m}\left |\sum_n\lambda_{\bm G} \frac{\braket{\Psi_{\bm q +\bm G}^{(0)}|\bar{\rho}_{\bm q}|\Psi_{\bm G,n}^{(0)}}\braket{\Psi_{\bm G,n}^{(0)}|\bar{\rho}_{\bm G}|\Psi_0^{(0)}}}{E_0^{(0)}-E_{\bm G,n}^{(0)}}\right |^2 \delta[\epsilon-(E_{\bm q+\bm G,m}^{(0)}-E_0^{(0)})].
\end{equation}
where we have not included the excitation energy shifts which are formally of higher order in $\lambda_{\bm G}$. Next, we employ the SMA $\ket{\Psi_{\bm q}}=\frac{1}{\sqrt{\,\overline{S}_{\bm q}\,N}  }\bar{\rho}_{\bm q}\ket{\Psi_0}$ so that
\begin{equation}
    S(\bm q,\epsilon)=\frac{1}{N} \sum_{\bm G} \left |\lambda_{\bm G}\frac{ \braket{\Psi_{\bm q+\bm G}|\bar{\rho}_{\bm q}|\Psi_{\bm G}} \braket{\Psi_{\bm G}|\bar{\rho}_{\bm G}|\Psi_{0}^{(0)}} }{\Delta_{\bm G}}\right |^2 \delta{(\epsilon-\Delta_{\bm q+ \bm G})},
    \label{eq:SM_SMA_S}
\end{equation}
and noting that the second expectation value in the numerator $\braket{\Psi_{\bm G}|\bar{\rho}_{\bm G}|\Psi_{0}^{(0)}}=\sqrt{N \bar{S}_{\bm G}}$, \cref{eq:SM_SMA_S} becomes
\begin{equation}
    S(\bm q,\epsilon)=\sum_{\bm G} \left |\lambda_{\bm G} \frac{\braket{\Psi_0^{(0)}|\bar{\rho}_{\bm q+\bm G} ~\bar{\rho}_{\bm q} ~\bar{\rho}_{\bm G} | \Psi_{0}^{(0)}}  }{N \sqrt{\bar{S}_{\bm q+\bm G} } \,\Delta_{\bm G}} \right |^2 \delta(\epsilon -\Delta_{\bm q+\bm G}).
\end{equation}
We now examine the small $\bm q$ limit of the three point function $\braket{ \overline{\rho}_{\bm{q}+\bm{G}}^{\dagger} \overline{\rho}_{\bm{q}} \overline{\rho}_{\bm{G}}}_0$, see also \cite{Fertig_QGDipole}. Since there are no dipole allowed transitions within the 
lowest Landau level $\overline{\rho}_{\bm{q}}|\Psi_0\rangle = 0$ to first order in $|\bm{q}|$. Therefore, we can replace $\overline{\rho}_{\bm{q}} \overline{\rho}_{\bm{G}}$ by the commutator $[\overline{\rho}_{\bm{q}},\overline{\rho}_{\bm{G}}]$(since the second term will vanish by the argument above in the long wavelength limit). Using the GMP algebra,
\begin{equation}
\begin{aligned}
    \braket{ \overline{\rho}_{\bm{q}+\bm{G}}^{\dagger} \overline{\rho}_{\bm{q}} \overline{\rho}_{\bm{G}}}_0 &=\braket{ \overline{\rho}_{\bm{q}+\bm{G}}^{\dagger} [\overline{\rho}_{\bm{q}},\overline{\rho}_{\bm{G}}]}_0= (e^{q^* G \ell^2/2}-e^{G^*q \ell^2/2}) \braket{\overline{\rho}_{\bm{q}+\bm{G}}^{\dagger}\overline{\rho}_{\bm q+\bm G}}_0 \\ &=N (e^{q^* G \ell^2/2}-e^{G^*q \ell^2/2}) \overline{S}_{\bm q+\bm G}
\end{aligned}
\end{equation}
Expanding the exponentials to first order in $q$:
\begin{equation}
    (e^{q^* G \ell^2/2}-e^{G^*q \ell^2/2})=1+\frac{q^* G \ell ^2}{2}-1-\frac{G^*q \ell ^2}{2}+O(q^2)=i \ell ^2 (\bm q \times\bm G)_{\hat{z}}+O(q^2),
\end{equation}
so to leading order in $q$
\begin{equation}
    \braket{ \overline{\rho}_{\bm{q}+\bm{G}}^{\dagger} \overline{\rho}_{\bm{q}} \overline{\rho}_{\bm{G}}}_0/N= i \ell ^2 (\bm q \times\bm G)_{\hat{z}} \overline{S}_{\bm q+\bm G}.
\label{eq:small_q_3pt}
\end{equation}
Next, we compare this expression to the MC simulations result. In Fig. \ref{fig:comparison} we plot the left-hand-side of Eq. \cref{eq:small_q_3pt} obtained from the MC simulations and the long-wavelength limit (right-hand-side of Eq. \cref{eq:small_q_3pt} for a representative $\bm{G}= |\bm{G}|(-1,0)$. Fig. \cref{fig:comparison} shows the two expressions agree for small $\bm q$ and provides a range of validity for this long wavelength limit. 

\subsection{Higher magnetoroton bands and exact diagonalization results}
\begin{figure}
    \centering   \includegraphics[width=0.8\textwidth]{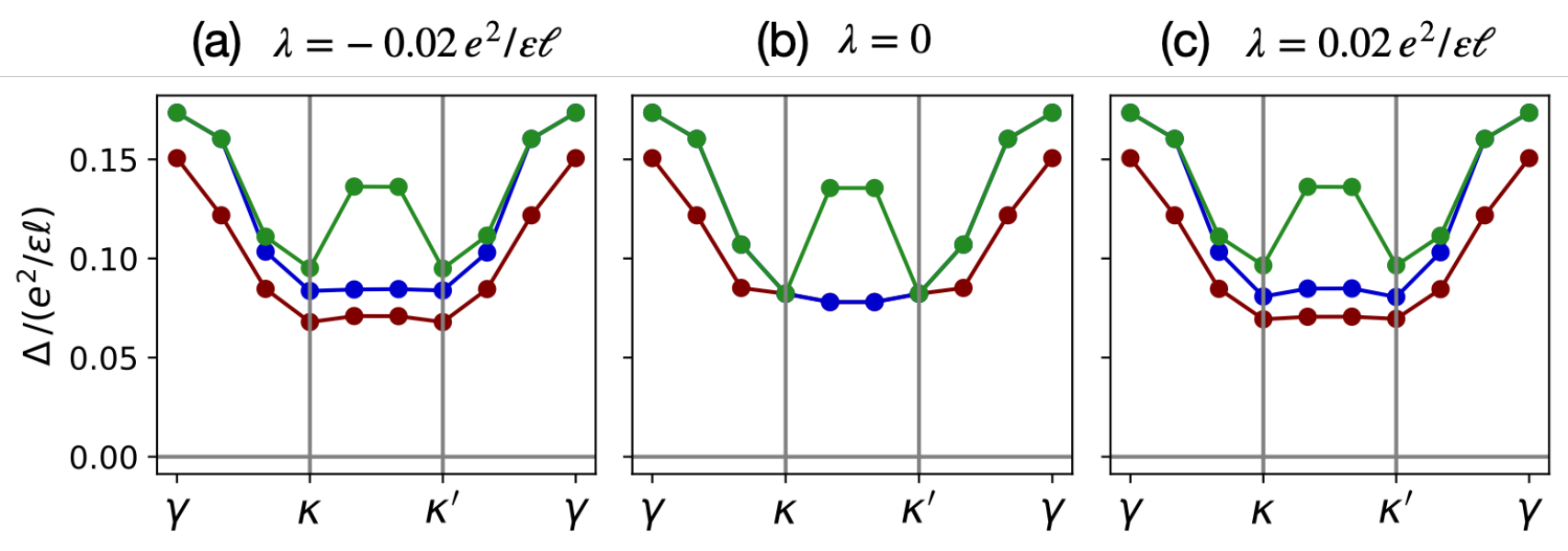}
    \caption{Three lowest magnetoroton bands for three values of the potential strength and for one flux quantum per unit cell. The path in the Brillouin zone is the same as the one indicated in \cref{fig:MR_Bandstructure}(a) in the main text. The $\bm{\kappa}/\bm{\kappa}^{\prime}$ points where the magnetoroton softening occurs are indicated by vertical lines.}
    \label{fig:Other_Bands}
\end{figure}
\begin{figure}
    \centering   \includegraphics[width=0.85\textwidth]{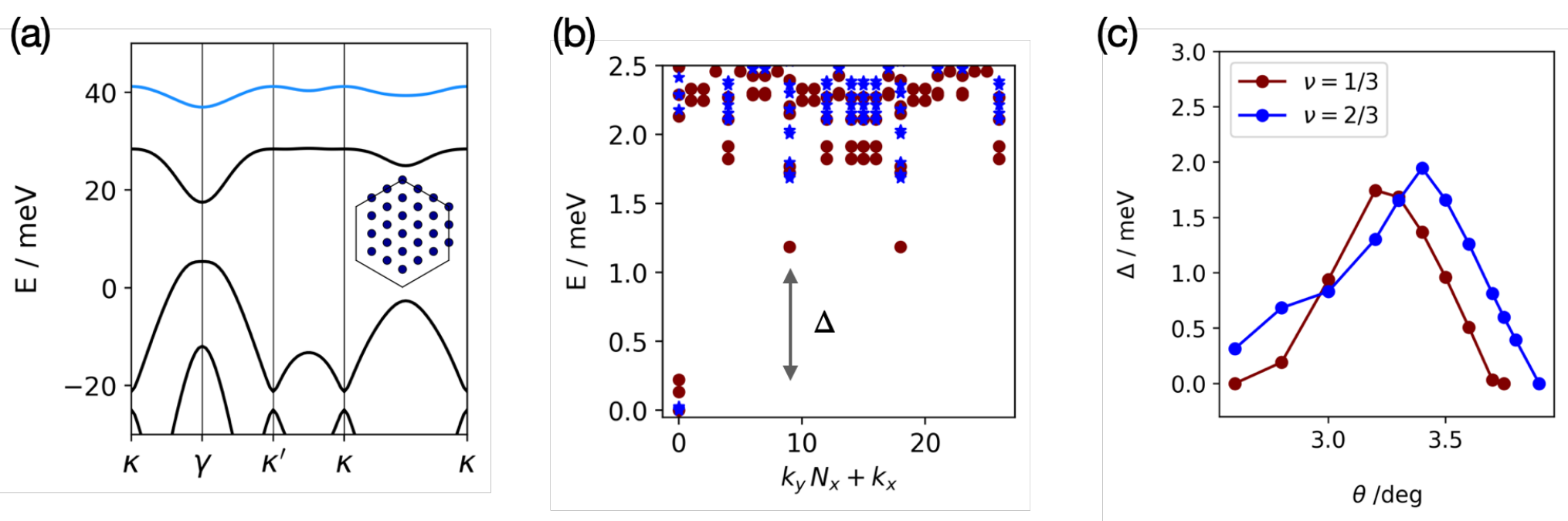}
    \caption{(a) Moiré band structure for the continuum model for MoTe$_2$ obtained in \cite{FCI_DiXiao}, for $\theta=3.5^{\circ}$. We project Coulomb interactions to the topmost band and obtain the many-body spectrum via exact diagonalization for a system with 27 unit cells, which corresponds to the discretization od the Brillouin zone indicated in the inset. (b) ED spectrum for $\nu=1/3$ (red) and $\nu=2/3$ (blue) at $\theta=3.5^{\circ}$ and for $\varepsilon=20$. We consider a valley-polarized ground state. (c) Many-body gap $\Delta$ for $\nu=1/3$ and $\nu=2/3$ as a function of twist angle.}
    \label{fig:ED_MoTe2}
\end{figure}

In \cref{fig:Other_Bands} we show the three lowest magnetoroton bands, obtained from diagonalizing \cref{eq:matrix_elements} in the main text, for the case of one flux quantum per unit cell. In the absence of the potential, $\lambda=0$, there is a three-fold degeneracy at ${\bm \kappa}/{\bm \kappa}'$, and a two-fold degeneracy across the line between ${\bm \kappa}$ and ${\bm \kappa}'$, as expected. When the potential is turned-on, the three states at ${\bm \kappa}/{\bm \kappa}'$ split, as do the two across the ${\bm \kappa}$-to-${\bm \kappa}'$ line. For $\lambda=0$ the magnetoroton minimum is located in the proximity of the $\bm{m}$-point, once the hexagonal potential is introduced the magnetoroton minimum shifts to ${\bm \kappa}/{\bm \kappa}'$, due to stronger level repulsion between the three coupled states.\\\\
%We observe that for $|\lambda|>0.01 e^2/\varepsilon \ell$ the third magnetoroton band already merges into the continuum of particle-hole excitations, which is not shown in the figures.
In order to make a connection between our results for the magnetoroton gap and the corresponding quantity in realistic materials, we perform exact diagonalization (ED) on a continuum model for tMoTe$_2$ with parameters $V=20.8$ meV, $\psi=107.7^{\circ}$ and $\omega=-23.8$ meV \cite{FCI_DiXiao}. We project Coulomb interactions with $\varepsilon=20$ to the topmost moiré band and obtain the many-body spectrum for a system with 27 unit cells, for details on the ED method for continuum models see \cite{FCI_Flatiron,FCI_DiXiao,FCI_LiangFu}. \cref{fig:ED_MoTe2}(a) shows a typical band structure obtained from the continuum model of MoTe$_2$ and in \cref{fig:ED_MoTe2}(b) we present typical many-body spectra for both $\nu=1/3$ and $\nu=2/3$ fillings. We focus on the valley-polarized sector and consider twist angles where the ground state is an FCI. From the spectrum we calculate the neutral gap $\Delta$ as the difference between the lowest excitation and the highest of the three quasi-degenerate topological groundstates, and plot it in \cref{fig:ED_MoTe2}(c) for both fillings. Using that the periodic potential encloses on flux quantum per unit cell we obtain a relation between the magnetic length and the moiré length $\ell=(\sqrt{3}/4\pi)^{1/2}a_M$, which we can use to convert from $e^2/\varepsilon \ell$ to meV. The values for $\Delta$ reported in the main text for MoTe$_2$ correspond to the maximum value attained in \cref{fig:ED_MoTe2}(c) for each filling.

\end{document}